\documentclass[12pt]{article}
\usepackage{amsmath,amssymb,amsfonts,amsthm}

\title{Dynamics of entanglement of a three-level atom in motion interacting with two coupled modes including parametric down conversion}

\author{M J Faghihi$^{1,2,3}$, M K Tavassoly$^{1,2,*}$ and M Hatami$^{4}$ \\
 \footnotesize{$^1$ Atomic and Molecular Group, Faculty of Physics, Yazd University, Yazd, Iran} \\
 \footnotesize{$^2$ The Laboratory of Quantum Information Processing, Yazd University, Yazd, Iran} \\
 \footnotesize{$^3$ Physics and Photonics Department, Graduate University of Advanced Technology, Mahan, Kerman, Iran} \\
 \footnotesize{$^4$ Department of Physics, Faculty of Science, Shiraz University of Technology, Shiraz, Iran} \\
 \footnotesize{$^*$ E-mail: mktavassoly@yazd.ac.ir}}

\begin{document}
\maketitle

 \newcommand{\norm}[1]{\left\Vert#1\right\Vert}
 \newcommand{\abs}[1]{\left\vert#1\right\vert}
 \newcommand{\set}[1]{\left\{#1\right\}}
 \newcommand{\R}{\mathbb R}
 \newcommand{\I}{\mathbb{I}}
 \newcommand{\C}{\mathbb C}
 \newcommand{\eps}{\varepsilon}
 \newcommand{\To}{\longrightarrow}
 \newcommand{\BX}{\mathbf{B}(X)}
 \newcommand{\HH}{\mathfrak{H}}
 \newcommand{\A}{\mathcal{A}}
 \newcommand{\D}{\mathcal{D}}
 \newcommand{\N}{\mathcal{N}}
 \newcommand{\x}{\mathcal{x}}
 \newcommand{\p}{\mathcal{p}}
 \newcommand{\la}{\lambda}
 \newcommand{\af}{a^{ }_F}
 \newcommand{\afd}{a^\dag_F}
 \newcommand{\afy}{a^{ }_{F^{-1}}}
 \newcommand{\afdy}{a^\dag_{F^{-1}}}
 \newcommand{\fn}{\phi^{ }_n}
 \newcommand{\HD}{\hat{\mathcal{H}}}
 \newcommand{\HDD}{\mathcal{H}}

 \begin{abstract}
 In this paper, a model by which we study the interaction between a motional three-level atom and two-mode field injected simultaneously in a bichromatic cavity is considered; the three-level atom is assumed to be in a $\Lambda$-type configuration. As a result, the atom-field and the field-field interaction (parametric down conversion) will be appeared. It is shown that, by applying a canonical transformation, the introduced model can be reduced to a well-known form of the generalized Jaynes-Cummings model. Under particular initial conditions, which may be prepared for the atom and the field, the time evolution of state vector of the entire system is analytically evaluated. Then, the dynamics of atom by considering `atomic population inversion' and two different measures of entanglement, i.e., `von Neumann entropy' and `idempotency defect' is discussed, in detail. It is deduced from the numerical results that, the duration and the maximum amount of the considered physical quantities can be suitably tuned by selecting the proper field-mode structure parameter $p$ and the detuning parameters.
 \end{abstract}


  \section{Introduction}\label{sec-intro}
 Entanglement, an unbreakable quantum correlation between parts of a multipartite system, is one of the most essential characteristics of the quantum mechanical systems which plays a key role within new information technologies \cite{ benenti}. Also, it is an important resource in many interesting applications in fields related to quantum computation as well as quantum information \cite{bennett}. However, the appearance of entanglement in the interaction between light and matter in a cavity QED is of special interest, in which the atom-field interaction produces the entangled state. \\
 There is a fully quantum mechanical and of course an exact (in the rotating wave approximation) model that describes a two-level atom interacting with a single-mode field, as a very simplified version of atom-field interaction, which is called the Jaynes-Cummings model (JCM) \cite{JCM1,JCM2}.
 Many generalizations have been proposed to modify the JCM in the literature \cite{GJCM1,GJCM2,GJCM3,GJCM4,GJCM5,GJCM6,GJCM7,GJCM8,GJCM9}. In addition, various researches have been published to quantify the atom-field entangled states, using the standard JCM (and also the generalized JCM). For instance, a general formalism for a $\Lambda$-type three-level atom interacting with a correlated two-mode field is presented by Abdel-Aty {\it et al} \cite{aty}. The authors found the degree of entanglement (DEM) for their system by obtaining the density matrix operator. The entanglement properties of a cavity field generated from a laser-driven collective three-level atomic ensemble (in a $V$-type configuration) inside a two-mode cavity including the spontaneously generated coherence have been investigated by Tang {\it et al} \cite{tang}. Entanglement dynamics, as measured by concurrence, of the tripartite system of one atom and the two cavity modes has been discussed by Abdel-Aty {\it et al} \cite{abdel-aty}. Dynamics of entanglement and other nonclassical properties of a $V$- and a $\Lambda$-type three-level atom interacting with a single-mode field in a Kerr medium with intensity-dependent coupling and in the presence of the detuning parameters have been studied in \cite{zait} and \cite{us} by Zait and us, respectively. In particular, we illustrated that the strength and time interval of nonclassicality aspects is more visible in $\Lambda$-type than $V$-type three-level atoms \cite{us}. Recently, entanglement dynamics of the nonlinear interaction between a three-level atom ( in a $\Lambda$ configuration) and a two-mode cavity field in the presence of a cross-Kerr medium and its deformed counterpart \cite{newhonarasa}, intensity-dependent atom-field coupling and the detuning parameters has been discussed by us \cite{usJOSA,usJOSA1}. \\
 On one hand, various generalizations to modify the JCM have been proposed in the literature. For instance, one may consider a two-mode field simultaneously inject within a high-Q two-mode bichromatic cavity \cite{abd-AOP} so that the atom interacts with each field individually as well as both fields. As a result, other nonlinearities may be occurred during the atom-field interaction, for example, one may refer to switching, modulation and frequency selection of radiation in optical communication networks \cite{net}. One of the main goals of this research is to investigate the effect of field-field interaction, namely, parametric down conversion, on the atomic population inversion as well as the entanglement dynamics between subsystems (atom and field).
 On the other hand, due to the fact that in any atom-field interaction, the atom may not be exactly static during the interaction, the influence of atomic motion on the interaction dynamics should be taken into account. For instance, this effect together with field-mode structure on the atomic dynamics (atomic population inversion) has been examined by Joshi {\it et al} \cite{joshi1}. Also, a model in which a moving atom undergoes a two-photon transition in a two-mode coherent state field has been studied by Joshi \cite{joshi2}. The authors then compared their own results with those of \cite{schlicher} by Schlicher in which an atom undergoes a one-photon transition. As a further comment about the realization of atomic motion, note that, there exist some experiments that are comparable to the interaction of an atom with an electromagnetic pulse \cite{experiment1,experiment2}, in which the interaction of an atom with cavity eigenmodes of different shape functions is studied. \\
 In this paper, we try to present a model that describes a moving three-level atom (in a $\Lambda$-configuration) interacting with a two-mode field injected simultaneously in a bichromatic cavity in the presence of the detuning parameters.
 Apart from other new features of our work, in particular, we investigate individually and simultaneously the effects of atomic motion (via varying the field-mode structure parameter) and the detuning parameters on some physical criteria which will be studied in detail.
 Indeed, the main goal of the present paper is to discuss on the effects of these parameters on the physical appearances of the state vector of the whole system. To achieve this purpose, we examine the dynamics of atomic population inversion, the time evolution of the field entropy by which the amount of the degree of entanglement (DEM) between subsystems is determined and the decoherence (coherence loss) which is obtained by idempotency defect.\\
 The remainder of paper is organized as follows: In the next section, we obtain the state vector of the whole system using the generalized JCM. In section 3, the atomic dynamics is discussed by considering atomic population inversion. Then, using the von Neumann approach,  the time evolution of the field entropy is studied in sections 4, and section 5 deals with the decoherence by considering the linear entropy. Finally, section 6 contains a summary and concluding remarks.

 \section{Introducing the model and its solution}
This section is devoted to describe the interaction between a moving three-level atom and two coupled modes of the cavity field, taking into account the field-field interaction by considering parametric down conversion. So, let us assume a model in which the two-mode quantized electromagnetic field oscillating with frequencies $\Omega_{1}$ and $\Omega_{2}$ in an optical cavity interacts with the $\Lambda$-type three-level atom which is free to move in the cavity. In this atomic configuration, the atomic levels are indicated by $|j\rangle$ with energies $\omega_{j}$, where $j=1,2,3$, the transitions $|1\rangle\rightarrow|2\rangle$ and $|1\rangle\rightarrow|3\rangle$ are allowed and the transition $|2\rangle\rightarrow|3\rangle$ is forbidden in the electric-dipole approximation \cite{zubairy}. Anyway, the Hamiltonian describing the dynamics of our above system in the RWA can be written as ($\hbar = c = 1$):
 \begin{eqnarray}\label{H}
 \hat{H} =\hat{H}_{A}+\hat{H}_{F}+ \hat{H}_{AF} + \hat{H}_{FF},
 \end{eqnarray}
 where the atomic and field parts of the Hamiltonian read as
 \begin{eqnarray}\label{H-PA}
 \hat{H}_{A}= \sum_{j=1}^{3} \omega_{j}\hat{\sigma}_{jj},   \;\;\;\;\;   \hat{H}_{F}= \sum_{j=1}^{2} \Omega_{j} \hat{a}^{\dag}_{j} \hat{a}_{j},
 \end{eqnarray}
 and the atom-field and field-field interactions are given
 \begin{eqnarray}\label{H-PB}
 \hat{H}_{AF} &=& \sum_{j=1}^{2} \left[ g_{1}^{(j)}f_{1}^{(j)}(z)(\hat{a}_{j}\; \hat{\sigma}_{12}+\hat{\sigma}_{21}\hat{a}_{j}^{\dag})
 + g_{2}^{(j)}f_{2}^{(j)}(z)(\hat{a}_{j}\;\hat{\sigma}_{13}+\hat{\sigma}_{31}\hat{a}_{j}^{\dag}) \right], \nonumber \\
 \hat{H}_{FF} &=& \mathbf{g} \left( \hat{a}_{1}^{\dag}\hat{a}_{2} +  \hat{a}_{1}\hat{a}_{2}^{\dag}  \right),
 \end{eqnarray}
 where $\hat{\sigma}_{ij}$ is the atomic ladder operator between the levels $|i\rangle$ and $|j\rangle$ defined by $\hat{\sigma}_{ij}=|i\rangle \langle j|,(i,j=1,2,3),\hat{a}_{j}$ ($\hat{a}_{j}^{\dag}$) is the bosonic annihilation (creation) operator of the field mode $j$, the constants $g_{1}^{(j)}, g_{2}^{(j)}$ determine the strength of the atom-field couplings for the mode $j$, and $\mathbf{g}$ denotes the field-field coupling constant. Notice that, the influence of atomic motion in the model has been entered by the shape functions $f_{1}^{(j)}(z)$ and $f_{2}^{(j)}(z)$. \\
 It is worth to note that a deep view in the Hamiltonian of the atom-field interaction implies the fact that this Hamiltonian may be reconstructed by changing $g_{i}^{(j)},(i,j=1,2),$ to $g_{i}^{(j)}f_{i}^{(j)}(z)$, when it is compared with the JCM where the atomic motion is neglected, i.e. in our model, the atom-field coupling depends on the atomic motion by considering the shape function $f_{i}^{(j)}(z)$. \\
 Now, in order to obtain the solution of the atom-field system described by the above Hamiltonians, there are three different but equivalent methods, namely probability amplitudes, Heisenberg operators and the unitary time evolution operator approaches \cite{zubairy}. However, the presented formalism for the considered system is based on the method of time evolution operator. But, before using this approach to reach our goal, it is necessary to introduce the canonical transformations
 \begin{eqnarray}\label{Can-Trans}
 \hat{a}_{1} = \hat{b}_{1} \cos \theta + \hat{b}_{2} \sin \theta,\;\;\;\;\;\hat{a}_{2} = \hat{b}_{2} \cos \theta - \hat{b}_{1} \sin \theta,
 \end{eqnarray}
 which is the well-known Bogoliubov-Valatin transformation \cite{BVT1,BVT2,BVT3} and has been introduced in the context of the Bardeen-Cooper-Schrieffer model of superconductivity \cite{HBVT}. In this transformation, the operators $\hat{b}_{i} (\hat{b}_{i}^{\dag}), i=1,2,$ have the same meaning of the operators $\hat{a}_{i} (\hat{a}_{i}^{\dag})$ while $\theta$ is the rotation angle which will be determined later. It is worthwhile to mention that under these transformations, the sum of the photon number of the field is invariant, that is, $\hat{a}_{1}^{\dag} \hat{a}_{1} + \hat{a}_{2}^{\dag} \hat{a}_{2} = \hat{b}_{1}^{\dag} \hat{b}_{1} + \hat{b}_{2}^{\dag} \hat{b}_{2}$.\\
 Inserting the canonical transformations in (\ref{Can-Trans}) into the whole Hamiltonian in (\ref{H}) leads us to the following Hamiltonian
\begin{eqnarray}\label{H-F}
 \hat{\mathcal{H}} = \hat{H}_{0} + \hat{H}_{1},
 \end{eqnarray}
 where
 \begin{eqnarray}\label{H-0}
 \hat{H}_{0} =  \sum_{j=1}^{3} \omega_{j}\hat{\sigma}_{jj} + \sum_{j=1}^{2} \mathbf{\Omega}_{j} \hat{b}^{\dag}_{j} \hat{b}_{j},
 \end{eqnarray}
 and
 \begin{eqnarray}\label{H-1}
 \hat{H}_{1} =  \sum_{j=1}^{2} \left[ \mu_{1}^{(j)}f_{1}^{(j)}(z)(\hat{b}_{j}\; \hat{\sigma}_{12}+\hat{\sigma}_{21}\hat{b}_{j}^{\dag})
 + \mu_{2}^{(j)}f_{2}^{(j)}(z)(\hat{b}_{j}\;\hat{\sigma}_{13}+\hat{\sigma}_{31}\hat{b}_{j}^{\dag}) \right],
 \end{eqnarray}
 with
 \begin{eqnarray}\label{H-Coeff}
 \mathbf{\Omega}_{1} &=& \Omega_{1} \cos^{2} \theta + \Omega_{2} \sin^{2} \theta - \mathbf{g} \sin 2 \theta, \nonumber \\
 \mathbf{\Omega}_{2} &=& \Omega_{1} \sin^{2} \theta + \Omega_{2} \cos^{2} \theta + \mathbf{g} \sin 2 \theta, \nonumber \\
 \mu_{k}^{(1)} &=& g_{k}^{(1)} \cos \theta - g_{k}^{(2)} \sin \theta, \nonumber \\
 \mu_{k}^{(2)} &=& g_{k}^{(1)} \sin \theta + g_{k}^{(2)} \cos \theta, \;\;\;  k=1,2,
 \end{eqnarray}
 in which the rotation angle $\theta$ is still unknown and must be determined. To attain this aim, the evanescent wave terms from the Hamiltonians related to the field and field-field interaction should be avoided. Therefore, one may set the particular choice of angle $\theta$ which reads as
 \begin{eqnarray}\label{r}
 \theta = \frac{1}{2} \tan^{-1} \left( \frac{2  \mathbf{g}}{ \Omega_{2} - \Omega_{1} } \right).
 \end{eqnarray}
 With the above selection of $\theta$, the field-field coupling parameter $\mathbf{g}$ will then be in the form
 \begin{eqnarray}\label{g}
  \mathbf{g} = \frac{\delta \left( \Omega_{2} - \Omega_{1} \right) }{1 - \delta^{2}},
 \end{eqnarray}
 where $ g_{1}^{(1)}/g_{1}^{(2)} = \delta = g_{2}^{(1)}/g_{2}^{(2)}  $.
 Looking deeply at the relations (\ref{H-0}) and (\ref{H-1}) and comparing them with (\ref{H})-(\ref{H-PB}) shows clearly that, the applied canonical transformations simplify the interaction Hamiltonian by eliminating the field-field interaction. In this way, our presented model is reduced to the usual form of the JCM. \\
 Anyway, for next purpose, it is convenient to rewrite the Hamiltonian (\ref{H-F}) in the interaction picture; accordingly one may arrive at
 \begin{eqnarray}\label{VI}
  V_{I}(t) =  \mu f(z) \left[ \hat{b}_{2} \hat{\sigma}_{12} \exp (- i \Delta_{2} t ) +  \gamma \hat{b}_{2} \hat{\sigma}_{13} \exp (- i \Delta_{3} t )  \right] + \mathrm{c.c.},
 \end{eqnarray}
 with
 \begin{eqnarray}\label{VI-Coeff}
  \mu = g \sqrt{1 + \delta^{2}},\;\;\;  \gamma = \mu_{2}^{(2)}/ \mu_{1}^{(2)}, \;\;\;  \Delta_{k} = \mathbf{\Omega}_{2} - (\omega_{1} - \omega_{k}),\;\;\;k=2,3,
 \end{eqnarray}
 where we redefined $g_{1}^{(2)} = g$. In last step, we have to determine the explicit form of the functions $f_{j}(z)$, $j=1,2$, which correspond to the atomic motion. To achieve this purpose, we restrict the atomic motion in the $z$-axis direction which is consistent with respect to the cavity QED experiments. Therefore, only the $z$-dependence of the field-mode function would be necessarily taken into account and so the atomic motion may be specified as
 \begin{eqnarray}\label{flamb}
  f_{i}(z) \rightarrow f_{i}(vt),\;\;\;\;i=1,2,
 \end{eqnarray}
 where $v$ denotes the atomic velocity. To make the latter discussion more convenient, one may define a $\mathrm{TEM_{mnp_{i}}}$ mode as \cite{joshi1,joshi2,schlicher}
 \begin{eqnarray}\label{fz}
 f_{i}(z)=\sin(p_{i}\pi vt/L ),
 \end{eqnarray}
 where $p_{i}$ represents the number of half-wavelengths of the field mode inside a cavity with a length $L$. \\
 We can now find the explicit form of the wave function of the entire system by using the standard technique. So let us consider the initial state of the whole system to be in the following form:
 \begin{eqnarray}\label{sayi}
 |\psi(0)\rangle_{\mathrm{A-F}}=|1\rangle \otimes \sum_{n=0}^{+\infty} \sum_{m=0}^{+\infty} q_{n} q_{m} |n,m \rangle = \sum_{n = 0}^{+\infty} \sum_{m = 0}^{+\infty} q_{n} q_{m} |1, n,m \rangle,
 \end{eqnarray}
  where $q_{n}$ and $q_{m}$ are the probability amplitudes of the initial radiation field of the cavity. Keeping in mind all above assumptions in addition to considering $p_{1} =p_{2} \equiv p$ (without loss of generality), it may be found that, by the action of the time evolution operator directly (with the Hamiltonian in (\ref{VI})), on the initial state vector of the system in (\ref{sayi}), we arrive at the explicit form of the wave function as follows
 \begin{eqnarray}\label{say}
 \hspace{-2cm} |\psi(t)\rangle &=& \sum_{n=0}^{+\infty} \sum_{m=0}^{+\infty} q_{n} q_{m} \Big[ A(n,m,t) |1,n,m \rangle \nonumber \\
 &+& B(n,m+1,t) |2,n,m+1\rangle + C(n,m+1,t)|3,n,m+1\rangle \Big]
 \end{eqnarray}
 where $A,B$ and $C$ are the atomic probability amplitudes which may be evaluated by a lengthy but straightforward procedure as
 \begin{eqnarray}\label{saycoef}
 A(n,m,t)&=&\cos \left[ \sqrt{ \left( |\Theta_{1}|^{2} + |\Theta_{2}|^{2}   \right) (m+1) } \right], \nonumber \\
 B(n,m+1,t)&=& \frac{\Theta_{1}^{*}}{i \sqrt{ \left( |\Theta_{1}|^{2} + |\Theta_{2}|^{2}   \right) }} \sin \left[ \sqrt{ \left( |\Theta_{1}|^{2} + |\Theta_{2}|^{2}   \right) (m+1) } \right], \nonumber \\
 C(n,m+1,t)&=&  \frac{\Theta_{2}^{*}}{i \sqrt{ \left( |\Theta_{1}|^{2} + |\Theta_{2}|^{2}   \right) }} \sin \left[ \sqrt{ \left( |\Theta_{1}|^{2} + |\Theta_{2}|^{2}   \right) (m+1) } \right],
 \end{eqnarray}
 with the following definitions for $\Theta_{i}(t) (i=1,2)$
 \begin{eqnarray}\label{theta1}
 \Theta_{1}(t) &=& \mu \int_{0}^{t}f(vt') \exp(- i \Delta_{2} t' ) dt', \nonumber \\
 \Theta_{2}(t) &=& \gamma \mu \int_{0}^{t}f(vt') \exp(- i \Delta_{3} t' ) dt'.
 \end{eqnarray}
 For a particular choice of the atomic motion, we consider the velocity of the atom by the special value $v=g L/\pi$ and hence the equation (\ref{fz}) is reduced to $f(z)=\sin(p g t)$ (recall that we assumed $p_{1} = p_{2} = p$). Consequently, the exact forms of $\Theta_{i}(t)$ are as bellow:
 \begin{eqnarray}\label{thetaf}
 \Theta_{1}(t) &=& \mu \left[ \frac{\sin \left( \frac{ p - \Delta_{2} }{ 2 } \; t \right) }{i(p - \Delta_{2})} \exp \left( i \; \frac{ p - \Delta_{2} }{ 2 } \; t \right) - \frac{\sin \left( \frac{ p + \Delta_{2} }{ 2 } \; t \right) }{i(p + \Delta_{2})} \exp \left( - i \; \frac{ p + \Delta_{2} }{ 2 } \; t \right) \right], \nonumber \\
 \Theta_{2}(t) &=& \gamma \mu \left[ \frac{\sin \left( \frac{ p - \Delta_{3} }{ 2 } \; t \right) }{i(p - \Delta_{3})} \exp \left( i \; \frac{ p - \Delta_{3} }{ 2 } \; t \right) - \frac{\sin \left( \frac{ p + \Delta_{3} }{ 2 } \; t \right) }{i(p + \Delta_{3})} \exp \left( - i \; \frac{ p + \Delta_{3} }{ 2 } \; t \right) \right].
 \end{eqnarray}
 It is now necessary to emphasize the fact that the obtained state vector of the entire system in (\ref{say}) is authentic for arbitrary amplitudes of the initial states of the field such as number, phase, coherent or squeezed state. However, since the coherent state is more accessible than other typical field states (recall that the laser field far above the threshold condition is known as a coherent state \cite{zubairy}), we shall consider the fields to be initially in the coherent states
 \begin{eqnarray}\label{amplitude}
 |\alpha, \beta \rangle = \sum_{n=0}^{+\infty}  \sum_{m=0}^{+\infty} q_{n} q_{m} |n,m \rangle, \;\;\;\;q_{n} = \exp \left( -\frac{ |\alpha|^{2} }{2}\right) \frac{\alpha^{n}}{\sqrt{n!}}, \;q_{m} = \exp \left( -\frac{ |\beta|^{2} }{2}\right) \frac{\beta^{m}}{\sqrt{m!}},
 \end{eqnarray}
 in which $|\alpha|^{2}$ and $|\beta|^{2}$ denote the mean photon number (intensity of light) of mode $1$ and $2$, respectively. \\
 Inserting (\ref{thetaf}) in the time-dependent coefficients (\ref{saycoef}) leads one to the explicit form of the state vector and consequently the density matrix of the entire atom-field system.

 \section{Atomic population inversion}\label{S-inversion}
 We are now in a position to examine the atomic dynamics, in particular the dynamics of an important quantity, namely atomic population inversion. For the present model (including a `three-level' atom in the $\Lambda$-type configuration), the atomic inversion, which is introduced as the difference between the excited-state and ground-state probabilities, may be defined as follows \cite{Inv1,Inv2}:
 \begin{eqnarray}\label{inversion}
 \langle \hat{\sigma}_{z}(t) \rangle = \rho_{11}(t) - (\rho_{22}(t) + \rho_{33}(t)),
 \end{eqnarray}
 in which the matrix elements of atomic density operator are generally given by
 \begin{eqnarray}\label{rhoshekl}
 \rho_{i j}(t) = \sum_{n=0}^{+\infty} \sum_{m=0}^{+\infty} \langle n,m, i | \psi(t) \rangle \langle \psi(t) | n,m, j \rangle,\;\;\; i ,  j = 1,2,3.
 \end{eqnarray}
 Figure \ref{Ainversion} shows the evolution of the atomic population inversion against the scaled time $\tau = g t$ for initial mean number of photons fixed at $|\alpha|^{2} = 10=|\beta|^{2}$. The left plots of this figure show the influence of atomic motion by selecting the fixed value of the field-mode structure parameter $p = 2$. Also, the effect of this parameter by considering different values of $p$ in the shape function $f(z)$ is discussed in the right plots.
 In figure \ref{Ainversion}(a), the exact resonant case is assumed ($\Delta_{2} = \Delta_{3} =0$) and the coupling constants are equal ($\gamma = 1$).
 Figure \ref{Ainversion}(b) is plotted to indicate the effect of different atom-field couplings by considering  $\gamma = 2$.
 The effect of detuning parameters ($\Delta_{2}/g =7, \; \Delta_{3}/g = 15$) together with the situations in which the atom-field couplings are equal ($\gamma = 1$) and different ($\gamma = 2$) is depicted in figure \ref{Ainversion}(c) and \ref{Ainversion}(d), respectively. \\
 In detail, it is seen from the left plots that, in resonance case (figures \ref{Ainversion}(a) and \ref{Ainversion}(b)), the atomic inversion oscillates between its minimum ($ -1$) maximum ($+1$) values. The temporal behaviour of this quantity is changed when the effect of detuning parameters is regarded (figures \ref{Ainversion}(c) and \ref{Ainversion}(d)), in which the atomic population is observed for all the time.

 In the right plots, the effect of field-mode structure parameter is studied by choosing different values of $p$. Figures \ref{Ainversion}(a) and \ref{Ainversion}(b) indicate that, by an increase of $p$, the fluctuation between minima and maxima becomes faster. It is understood from figures \ref{Ainversion}(c) and \ref{Ainversion}(d) that, in spite of the fact that the atomic population inversion is seen at all times for fixed $p = 2$ (left plot), this quantity periodically varies between the negative and positive values (the atomic inversion is observed, but not always), when the value of $p$ goes up (right one).
 Finally, it is found from figures \ref{Ainversion}(b) and \ref{Ainversion}(d) that, the ratio of atom-field coupling constants ($\gamma$) has no considerable effect on the amount of the population inversion.

 \section{Quantum mutual information and the DEM}
  In order to study the dynamics of entanglement and obtain the DEM, the quantum mutual information (quantum entropy) is a useful quantity that lead us to the amount of entanglement \cite{chuang}. In other words, temporal behaviour of quantum entropy of subsystems, here atom and field, indicates the time evolution of the DEM between them. In the present formalism, we use the von Neumann entropy \cite{pk1}. Before this, recalling the important theorem  by Araki and Leib \cite{araki} should be valuable. Based on this theorem, in a bipartite quantum system, the system and subsystem entropies, at any time $t$, are confined by the following triangle inequality
 \begin{eqnarray}\label{valen}
 |S_{A}(t)-S_{F}(t)|\leq S_{AF}(t) \leq S_{A}(t)+S_{F}(t),
 \end{eqnarray}
 where $S_{A}$, $S_{F}$ and $S_{AF}$ refer to the atom, the field and the total entropy, respectively. As a result, being the initial state of the whole system at a pure state implies the fact that the total entropy of the system is zero and remains constant. In other words, if a system initially is prepared in a pure state, the time evolution of the reduced entropies of subsystems is exactly identical. This fact means that for our consideration, $S_{A}(t)=S_{F}(t)$  at any time $t$ \cite{phoenix}. Therefore, by calculating the reduced entropy of the atom, the DEM will be obtained. \\
 Anyway, according to the von Neumann entropy, as a measure of entanglement, the entropy of the atom is defined by
 \begin{eqnarray}\label{ventd}
 S_{A}(t)=-\mathrm{Tr}_{A} \left(\hat{\rho}_{A}(t) \ln \hat{\rho}_{A}(t) \right) = S_{F}(t),
 \end{eqnarray}
 in which $\hat{\rho}_{A}(t)=\mathrm{Tr}_{F}\left(  |\psi(t) \rangle \langle \psi(t) |  \right)$ where $|\psi(t) \rangle$ has been introduced in (\ref{say}).
 Following the same procedure of \cite{usJOSA}, the entropy of the field (and so the atom, too) can be expressed by
 relation \cite{us,chuang,pk3}
 \begin{eqnarray}\label{sff}
 \mathrm{DEM}(t) = S_{F}(t) = - \sum_{j=1}^{3}\zeta_{j} \ln \zeta_{j},
 \end{eqnarray}
 where $\zeta_{j}$, the eigenvalues of the reduced atomic density matrix, are given by Cardano's method as \cite{kardan}
 \begin{eqnarray}\label{ventkardan}
 \zeta_{j}&=&-\frac{1}{3}\xi_{1}+\frac{2}{3}\sqrt{\xi_{1}^{2}-3\xi_{2}}\cos\left[\varrho+\frac{2}{3}(j-1)\pi \right],\;\;\;j=1,2,3, \nonumber \\
 \varrho &=& \frac{1}{3}\cos^{-1}\left[ \frac{9\xi_{1}\xi_{2}-2\xi_{1}^{3}-27\xi_{3}}{2(\xi_{1}^{2}-3\xi_{2})^{3/2}}\right],
 \end{eqnarray}
 with
 \begin{eqnarray}\label{vzal}
 \xi_{1} &=& -\rho_{11}-\rho_{22}-\rho_{33}, \nonumber \\
 \xi_{2} &=& \rho_{11}\rho_{22}+\rho_{22}\rho_{33}+\rho_{33}\rho_{11} -\rho_{12}\rho_{21}-\rho_{23}\rho_{32}-\rho_{31}\rho_{13}, \nonumber \\
 \xi_{3} &=& -\rho_{11}\rho_{22}\rho_{33}-\rho_{12}\rho_{23}\rho_{31}-\rho_{13}\rho_{32}\rho_{21} +\rho_{11}\rho_{23}\rho_{32}+\rho_{22}\rho_{31}\rho_{13}+\rho_{33}\rho_{12}\rho_{21},
 \end{eqnarray}
 where the above matrix elements have been introduced in (\ref{rhoshekl}).
 It is instructive to declare that since the parameter $\xi_{1}$ in (\ref{vzal}) precisely shows the trace of density matrix with minus sign, thus the exact value of this parameter is clearly equal to $-1$. Concerning equations (\ref{sff})-(\ref{vzal}), one is able to find the DEM between the atom and field. We would like to state the fact that getting zero value of the DEM in equation (\ref{sff}) ($S_{A(F)}=0$) means that the subsystems are unentangled (the system of atom-field is separable). \\
 Our presented results in figure \ref{dem} indicate the evolution of the field entropy against the scaled time $\tau$, from which the DEM is studied. In this figure, we have considered the parameters similar to figure \ref{Ainversion}.
 Comparing the left plots of figures \ref{dem}(a) and \ref{dem}(b) shows that changing the ratio of coupling constants ($\gamma$) from $1$ to $2$ has no remarkable effect on the variation of the field entropy in the absence of the detuning parameters.
 However, as is seen from figures \ref{dem}(c) and \ref{dem}(d), in the presence of the detuning ($\Delta_{2}/g =7, \; \Delta_{3}/g = 15$), the amount of the DEM is obviously diminished. Also, it is valuable to notice that, unlike the figure \ref{dem}(b), selecting appropriate values of atom-field coupling constants may improve the DEM.

 Paying attention to the right plot of figure \ref{dem}(a) indicates that, by an increase in $p$, the intervals of time of the entanglement between the atom and field, become shorter. In other words, the variations between maxima and minima values of the field entropy in large $p$ is faster than the small one.
 The right plot of figure \ref{dem}(b) shows that for the case $\gamma = 2$,  the amount of the DEM is decreased with increasing the parameter $p$.
 It is observed from figure \ref{dem}(c) that, by an increase of $p$, the amount of DEM is clearly enhanced when the detuning is present. In addition, comparing the right plots of figure \ref{dem}(c) and \ref{dem}(d) states that the amount of entanglement in the presence of the detuning can be enriched by changing $ \gamma $ from $1$ to $2$.

 \section{Linear entropy and coherence loss}
 A simple and direct measure to obtain the degree of decoherence (coherence loss) is called the "idempotency defect" \cite{idempotency1,idempotency2,idempotency3}.
 This quantity, which may be considered as the lowest order approximation to the von Neumann entropy, is a good measure to understand the purity loss of the state of a quantum system. This parameter can be easily measured by the linear entropy, defined as \cite{idempotency4}
 \begin{eqnarray}\label{LinEnt}
 S (\hat{\rho}) = \mathrm{Tr} \left[ \hat{\rho}_{A}(t) (1 - \hat{\rho}_{A}(t))  \right],
 \end{eqnarray}
 where $\hat{\rho}_{A}(t)$ is the atomic density operator. According to the latter equation, the linear entropy is zero for a pure state, that is, $\mathrm{Tr} \; \hat{\rho}_{A}(t) = \mathrm{Tr} \; \hat{\rho}_{A}^{2}(t)$.  Consequently, nonzero values of linear entropy then indicate the non-purity of the state of the system. Also, it is valuable to notice that maximally entanglement and consequently, the most mixed state is revealed when linear entropy gets the value $1$. \\
 In figure \ref{loss} we have plotted the idempotency defect, from which the coherence loss is studied, in terms of scaled time $\tau$ for different chosen parameters assumed in figure \ref{Ainversion}.
The left plots of figures \ref{loss}(a) and \ref{loss}(b) indicate that the $\gamma$ parameter (the ratio of the coupling constants) has no outstanding role on the coherence in resonance case. By entering the detuning parameters ($\Delta_{2}/g =7, \; \Delta_{3}/g = 15$), decoherence is rapidly attenuated (the left plots of figures \ref{loss}(c) and \ref{loss}(d)).

 We should now concentrate on the right plots, where the effect of field-mode structure parameter is investigated by considering different values of $p$. It is observed that, in resonance case, by an increase of $p$, decoherence and consequently the purity of the state of the system can be sharply descended (the right plots of figures \ref{loss}(a) and \ref{loss}(b)). Although, this behaviour is quite different with the situation that the detuning is present (figures \ref{loss}(c) and \ref{loss}(d)). In this case, by ascending $p$, the linear entropy is evidently increased.
%
 \section{Summary and conclusion}\label{summary}
%
 In this paper, we have proposed a model to discuss the interaction between a moving $\Lambda$-type three-level atom and a two-mode field injected simultaneously into the cavity in the presence of the detuning parameters. As a result of the presented model, we have considered the field-field interaction, from which the parametric down conversion was taken into account. Using a particular canonical transformation, an analytical and also exact solution for our interaction model has been presented and the explicit form of the state vector of the whole atom-field system has been obtained. Next, atomic dynamics by considering the atomic population inversion has been numerically investigated. Also, two different types of entanglement consist of von Neumann entropy (to study the DEM) and idempotency defect or linear entropy (to discuss the decoherence) have been studied. In each case, the effect of atomic motion via considering different values of field-mode structure parameter $p$, has been examined in the presence/absence of the detuning parameters with the equal/different values of the atom-field coupling constants. The main results of the paper are listed in what follows. \\

 $i)$ {\it Atomic population inversion:} It is found that the presence of the detuning parameters leads to the visibility of the atomic population inversion at all times. Adding the effect of field-mode structure parameter $p$, which is studied by choosing different values of $p$, implies the fact that the population inversion changes between the negative and positive values (the atomic inversion is appeared, not at all times), when the value of $p$ grows up. It is also worthwhile to mention that the ratio of atom-field coupling ($\gamma$) has no effective role in appearing the population inversion. \\

 $ii)$ {\it Dynamics of entanglement:} It is observed that  the effect of the detuning parameters is sharply pulled down the entanglement between the atom and the field. Although, by increasing the field-mode structure parameter $p$, the DEM clearly grows up. In addition, it may be useful to state that the amount of entanglement in the presence of the detuning can be improved by setting $\gamma = 2$ relative to $\gamma = 1$. \\

 $iii)$ {\it Coherence loss:} It is shown that the coherence between subsystems in resonance case can be enriched by changing the $\gamma$ parameter, i.e., the ratio of the coupling constants. Also, adding the effect of the detuning rapidly weakens the decoherence. Paying attention to the effect of field-mode structure parameter $p$, it is observed that in resonance case, by an increase of this parameter, decoherence and consequently the purity of the state of the system can be sharply descended. Although, in the presence of the detuning parameters, by increasing $p$ the decoherence is obviously ascended. \\

 Finally, in general, according to the obtained numerical results, the depth and domain of each of the considered properties can be appropriately controlled by choosing the suitable field-mode structure parameter $p$ and the detuning parameters, when initial states of the subsystems are fixed.\\



 \end{document}